\documentstyle[aps,prd,preprint,eqsecnum,epsf]{revtex}
\input epsf.tex

\newcommand{\be}{\begin{equation}}
\newcommand{\ee}{\end{equation}}
\newcommand{\ba}{\begin{eqnarray}}
\newcommand{\ea}{\end{eqnarray}}
\newcommand{\bas}{\begin{eqnarray*}}
\newcommand{\eas}{\end{eqnarray*}}

\newcommand{\tr}{ {\rm{Tr}}}
\newcommand{\R}{ {\cal{R}}}

\tightenlines

\begin{document}

\draft

\title{More on counterterms in the gravitational action and anomalies}
\author{Marika Taylor-Robinson
\thanks{email: M.M.Taylor-Robinson@damtp.cam.ac.uk}}
\address{Department of Applied Mathematics and
      Theoretical Physics, \\ Centre for Mathematical Sciences,
      \\ Wilberforce Road, Cambridge CB3 0WA, United Kingdom }
\date{\today}

\maketitle

\begin{abstract}
{
The addition of boundary counterterms to the gravitational action of
asymptotically anti-de Sitter spacetimes permits us to define the
partition function unambiguously, without background subtraction.
We show that the inclusion of p-form fields in the 
gravitational action requires the addition of further counterterms 
which we explicitly identify. We also relate logarithmic divergences 
in the action dependent on the matter fields
to anomalies in the dual conformal field theories. In particular we find
that the anomaly predicted for the correlator of the
stress energy tensor and two vector currents in four dimensions agrees
with that of the ${\cal{N}} = 4$ superconformal $SU(N)$
gauge theory. 
}
\end{abstract}


\section{Introduction}
\label{sec:intro}

The Maldacena conjecture \cite{Ma}, \cite{GKP}, \cite{W1}, \cite{A}
asserts that there is an equivalence between a gravitational theory in
a $(d+1)$-dimensional anti-de Sitter spacetime and a conformal field
theory in a $d$-dimensional spacetime which can in some sense be
viewed as the boundary of the higher dimensional spacetime. The
formulation of this correspondence is made precise by equating the
partition functions of the two theories
\be
{\cal{Z}}_{\rm{AdS}} [\Phi_i] = {\cal{Z}}_{\rm{cft}} [\Phi^{0}_i]. \label{ads}
\ee
In the supergravity theory, the fields $\Phi^{0}_i$ correspond to boundary
data for bulk fields $\Phi_i$ which propagate in the $(d+1)$-dimensional
spacetime. However, on the field theory side, these fields correspond
to external source currents coupled to various operators. 

An interesting consequence of the Maldacena conjecture
is the natural definition of the gravitational action for asymptotically
anti-de Sitter spacetimes without reference to a background \cite{BK},
\cite{KLS}. The
consideration of the gravitational action has a long history,
particularly in the context of black hole thermodynamics \cite{GH}. One
difficulty that has always plagued this approach is that the gravity
action diverges. The traditional approach to this problem is to use a
background subtraction whereby one compares the action of a spacetime
with that of a reference background, whose asymptotic geometry matches
that of the solution is some well-defined sense. However, this
approach breaks down when there is no appropriate or obvious
background. 

The AdS/CFT correspondence tells us that if, as we expect, the dual
conformal field theory has a finite partition function, then 
to make sense of (\ref{ads})
we must be able to remove the divergences of the gravitational action
without background subtraction. The framework for achieving this is by
defining local counterterms on the boundary \cite{W1}, \cite{LT},
\cite{HK}, \cite{FG}. Consider the Einstein action in $(d+1)$ dimensions
\be
I = - \frac{1}{16 \pi G_{d+1}} \int_{M} d^{d+1}x \sqrt{g}
( \R + d(d-1) l^2)  - \frac{1}{8 \pi G_{d+1}} \int_{N} \sqrt{\gamma} K
 \label{bulk}
\ee
where $G_{d+1}$ is the Newton constant and $\R$ is the Ricci scalar.
As usual a boundary term must be included for the equations of
motion to be well-defined \cite{GH}, 
with $K$ the trace of the extrinsic curvature of the $d$-dimensional
boundary $N$ embedded into the $(d+1)$-dimensional manifold $M$. Then
provided that the metric near the conformal boundary can be expanded
in the asymptotically anti-de Sitter form
\be
ds^2 = \frac{dx^2}{l^2 x^2} + \frac{1}{x^2} \gamma_{ij} dx^{i} dx^{j},
\label{asym}
\ee
where in the limit $x \rightarrow 0$ the metric $\gamma$ is
non-degenerate, we may remove divergent terms
in the action by the addition of a counterterm action dependent only
on $\gamma$ and its covariant derivatives of the form \cite{BK},
\cite{EJM} 
\ba
I_{\rm{ct}} &=& \frac{1}{8 \pi G_{d+1}} \int_{N} d^dx \sqrt{\gamma}
\lbrace (d-1) l + \frac{1}{2 (d-2) l} R(\gamma) + \frac{1}{2 l^3 (d-4)
    (d-2)^2} ( R_{ij}(\gamma) R^{ij}(\gamma) \label{ct} \\
&& \hspace{70mm} - \frac{d}{4 (d-1)}
  R(\gamma)^2) + ....\rbrace. \nonumber
\ea
$R(\gamma)$ and $R_{ij}(\gamma)$ are the Ricci scalar and the Ricci
tensor for the boundary metric respectively. Combined these
counterterms are sufficient to cancel divergences for $d \le 6$, with
several exceptions. Firstly, in even dimensions $d = 2n$ one has
logarithmic divergences in the partition function which can be related to
the Weyl anomalies in the dual conformal field theory \cite{HK}.  
Secondly, if the boundary metric becomes degenerate
one can no longer remove divergences by counterterm
regularisation \cite{MMT}; this is a manifestation of the fact that the dual
conformal field theory does not have a finite partition function in
the degenerate limit.  

The purpose of this
paper is to discuss a third case in which one
needs to consider regularisation more carefully. Much of the
discussion of the gravitational action
to date has concerned the case where the only boundary data in
(\ref{ads}) stems from the gravitational field. 
If there are matter fields on $M$ additional
counterterms may be needed to regulate the action. The addition of
scalar fields to the bulk action has been considered in several recent
papers \cite{NOO1}, \cite{NOO2},
but the focus of our consideration here will be p-form fields. 
We will also briefly discuss scalar
fields with potentials derived from maximal gauged supergravity theories 
in general dimensions. 

The motivation for our work is to complete the construction of the
theoretical tools required to investigate gravitational physics in
anti-de Sitter backgrounds. With
this in mind, we are particularly interested in whether one needs
counterterms to define the action for charged black brane solutions
such as those recently constructed in \cite{LM}, \cite{CS}, \cite{KS},
\cite{MMT2}. 
It turns out that for many of the gauged supergravity solutions
constructed so far, such as charged black hole solutions in five
dimensions \cite{BCS}, \cite{BCS1},
one does not need any further counterterms to (\ref{ct}).
However, one does need to include
further counterterms both for charged black holes in four dimensions
and more generally for magnetically charged branes in higher
dimensions. 

The related logarithmic terms that arise in the gravitational action 
when $d$ is even have an interpretation in terms
of anomalies arising from mixing conformal and other symmetries 
in the dual conformal field theories. Although
the aim of this paper is not to reproduce in detail the anomalies in the
dual conformal field theory we will discuss one particular case,
namely the anomaly in the correlator of the stress energy tensor and
two vector currents in four dimensions. 

The plan of this paper is as follows. In \S \ref{two} we summarise the
matter dependent counterterms that are found to be required. In \S
\ref{three} we consider the analysis of counterterms for $p$-form
fields. In \S \ref{four} we discuss scalar fields in the context of
gauged supergravity theories. In \S \ref{foura} we consider in more
detail gauged supergravity in seven dimensions. 
In \S \ref{five} we briefly consider the
related anomalies in the dual conformal field theories. 

\section{Counterterm regularisation of the partition function} \label{two}
\noindent

We will analyse here first a truncated action of a generic
gauged supergravity theory, such that 
the Einstein action (\ref{bulk}) is extended to include a scalar field $\phi$ 
and a $p$-form $F_p$ such that 
\be
I_{\rm{bulk}} = - \frac{1}{16 \pi G_{d+1}} \int_{M} d^{d+1}x \sqrt{g}
\left [ \R - \frac{1}{2} (\partial \phi)^2 + V(\phi) - \frac{1}{2p}
  e^{\alpha \phi} F_p^2 \right ]. \label{genac}
\ee
We will assume the potential $V(\phi)$ to be of the form arising in
$(d+1)$-dimensional maximal (or for d=5 the nearest to maximal that
is known) supergravity. This restriction includes most interesting
known solutions since the associated potentials  
fall into this category. Most of our analysis refers to $p=2$ since 
the main case of interest for $p > 2$ fields is seven-dimensional 
gauged supergravity 
and the latter does not admit a truncation of this type. We 
will consider the analysis for seven-dimensional supergravity 
separately in \S \ref{foura}.

It is perhaps useful to summarise here the results of the analysis of the
next sections. In addition to counterterms depending only on the
induced boundary metric (\ref{ct}) and logarithmic divergences
relating to the Weyl anomalies of the dual theories in $d=2n$ we find
the following. Restricting the scalar potential as above and as 
defined in more detail in \S \ref{four}, we find that
there is a scalar field divergence {\it only} when $d=3$, which can be
removed by a counterterm of the form 
\be
I_{\rm{ct}} = \frac{5l}{256 \pi G_4} \int d^3x \sqrt{\gamma} (\phi)^2.
\ee
For $p$-form fields, there will be no divergences for $d < 2p$ but in 
$d= 2p$ there will be a logarithmic divergence of the action
\be
I_{\rm{log}} = \frac{1}{32 \pi p G_{d+1} l} \ln \epsilon \int d^{2p}x
\sqrt{\gamma^{0}} (F^0_p)^2,
\ee
where $F^0_p$ is the induced field on the boundary, 
whilst for $d > 2p$ we must include the counterterm
\be
I_{\rm{div}} =  \frac{1}{64 \pi p^2 G_{d+1} l } 
\frac{(d-4p)}{(d-2p)} \int d^dx e^{\alpha \phi} (F_p)^2.
\ee
There will be additional terms in the anomaly for $d = 2p +2n$
depending on derivatives of $F_p$ and its coupling to the
curvature, and correspondingly further counterterms for $d > 2p + 2n$.
We derive these explicitly for $p=2$ (\ref{logf6}) in the context of
Einstein-Maxwell theories and, in \S \ref{foura}, discuss the absence
of these terms in the context of   
seven-dimensional gauged supergravity. 

\section{p-form fields} \label{three}
\noindent

Suppose that a $(d+1)$-dimensional manifold $M$ of negative curvature 
has a regular $d$-dimensional conformal boundary $N$ in these of
\cite{Pen}. Then 
in the neighbourhood of the boundary $N$ we will assume that the
the metric can be expressed in the form (\ref{asym}) with the induced
hypersurface metric $\gamma$ admitting the expansion 
\be
\gamma_{ij} = \gamma^{0}_{ij} + x^2 \gamma^{2}_{ij} + x^4
\gamma^{4}_{ij} + x^6 \gamma^{6}_{ij}.... \label{exp}
\ee
If $M$ is an Einstein manifold with negative cosmological constant,
then according to \cite{FG}, \cite{GL} such an expansion always exists. For
solutions of gauged supergravity theories with matter fields, 
demanding that this expansion is well-defined as $x
\rightarrow 0$ will impose conditions on 
the matter fields induced on $N$. Note that when $d$ is even
there will in general also be a logarithmic term $h^d$ appearing at order
$x^{d}$. However, it can be shown that $\tr[ (\gamma^{0})^{-1} h^d]$ 
vanishes identically; it will not
contribute to the action and can be neglected from here on. 

In what follows we will use repeatedly the Ricci tensor for the metric
(\ref{asym}) which has the following components
\ba
\R_{xx} &=& \frac{-d}{x^2} - \frac{1}{2} \lbrace \tr(\gamma^{-1}
\partial^{2}_{x} \gamma) - \frac{1}{x} \tr(\gamma^{-1} \partial_{x}
\gamma) - \frac{1}{2} \tr(\gamma^{-1} \partial_{x} \gamma \gamma^{-1}
\partial_{x} \gamma) \rbrace \nonumber \\
\R_{ij} &=& - \frac{d l^2 \gamma_{ij}}{x^2} - l^2 \lbrace \frac{1}{2}
\partial_{x}^2 \gamma - \frac{1}{2x} \partial_{x} \gamma - \frac{1}{2}
  (\partial_{x} \gamma) \gamma^{-1} (\partial_{x} \gamma)
+ \frac{1}{4} (\partial_{x} \gamma) \tr(\gamma^{-1} \partial_{x}
\gamma) \nonumber \\
&&  \hspace{30mm} + R(\gamma) l^{-2} - \frac{(d-2)}{2x} \partial_{x} \gamma -
\frac{1}{2x} \gamma \tr(\gamma^{-1} \partial_{x} \gamma) \rbrace_{ij}
  \\
\R_{xi} &=& \frac{1}{2} (\gamma^{-1})^{jk} \left [ \nabla_{i}
  \gamma_{jk,x} - \nabla_{k} \gamma_{ij,x} \right ], \nonumber
\ea
where $\nabla$ is the covariant derivative associated with $\gamma$. 
Let us include just a minimally coupled p-form field into the Einstein
action (we will consider the scalar field case separately in the next
section) so that 
\be
I_{\rm{bulk}} = - \frac{1}{16 \pi G_{d+1}} \int_{M} d^{d+1}x \sqrt{g} 
\left [ \R + d(d-1) l^2 - \frac{1}{2p} F_{p}^2 \right ], \label{pform}
\ee
where for the present we have ignored possible Chern-Simons terms.
This action is a consistent truncation of gauged supergravity theories
in $d < 6$ but not for $d=6$ itself. However, it is interesting to
consider cosmological Einstein-Maxwell theory in $d=6$ in its own
right and we shall do so
here. Gauged supergravity in seven dimensions is discussed in \S \ref{foura}.
The equations of motion derived from the action (\ref{pform}) are
\ba
\R_{mn} &=& - d l^2 g_{mn} + \frac{1}{2} F_{(p)m q_1..q_{p-1}} F_{(p)
  n}^{q_1..q_{p-1}}
+ \frac{(1-p)}{2 p (d-1)} F_p^2 g_{mn}; \nonumber \\
\R &=& -d (d+1) l^2 + \frac{(d+1-2p)}{2p (d-1)} F_{p}^2; \label{eqm} \\
d \ast F_{p} &=& 0; \hspace{5mm} d F_{p} = 0. \nonumber
\ea
Let us assume that in the vicinity of the conformal boundary the
$p$-form field can be expanded as a power series in $x$ as
\be
F_{p} = F^{0}_{p} + x dx \wedge A^{1}_{p-1} 
+ x^2 F^{2}_{p} + x^2 dx \wedge A^{2}_{p-1} + x^3 F^{3}_{p} .... \label{exp1}
\ee
where $G^{i}_{k}$ is a $k$-form dependent only on $x^i$. One can
justify this form for the expression retrospectively by demanding that
one can satisfy all field equations. For example, having
explicitly chosen the asymptotic form of the metric we can't have
terms in this expansion which diverge as $x \rightarrow 0$, as can be
seen by inspection of the Einstein equations. 

Now the key point is that to 
preserve the asymptotic form
of the metric we will have to restrict the leading order $p$-form contribution
to the Ricci scalar to be of order $x^{2p}$ or smaller. If it were any
larger, the leading order form of the metric would be changed. 
Since the bulk action includes a term of the form 
\be
I_{\rm{bulk}} \sim \int_{M} d^{d+1}x \sqrt{g} F_{p}^2,
\ee
there will be an induced infra-red 
divergence in the action only if $(d+1) > 2 p$. To justify why we can
neglect Chern-Simons terms in (\ref{eqm}) note that if we 
have terms in the action for $d = (3p-2)$ of the form
\be
I_{\rm{CS}} = \int_{M} F_{p} \wedge F_{p} \wedge A_{p-1},
\ee
then their magnitude is constrained by (\ref{exp1}) to be at least 
as small as $x^2$ as one takes the limit $x \rightarrow 0$. Hence
Chern-Simons terms will affect only finite terms in the action and can
be ignored here. 

To satisfy the closure property for the $p$-form
we will have to the implement the conditions
\be
d F^{0}_{p} = 0; \hspace{5mm} d A^{1}_{p-1} + 2 F^{2}_{p} = 0,
\hspace{5mm}  d A^{2}_{p-1} + 2 F^{3}_{p} = 0,
\ee
and so on. Expanding out the $p$-form equation of motion we find the
leading order conditions that 
\be
D^{0}_{i_0} (F^{0}_{p})^{i_{0}..i_{p-1}} 
= (d -2p) l^2 (A^{1}_{p-1})^{i_1..i_{p-1}}; \hspace{5mm} D^{0}_{i_{1}}
(A^{1}_{p-1})^{i_{1}..i_{p-1}} = 0, \label{feq}
\ee
where all indices are raised and lowered in the metric $\gamma^0$. The
first equation tells us that $A^1$ acts as a source for $F^0$ whilst
the second equation effectively picks out a gauge for $A^1$. Note that
if $F^0$ vanishes - or in another words, the field induced on the
boundary vanishes - then the field equations force the next order term
in the $p$-form to be at least of order $x^2$. 

\subsection{Vector fields}
\noindent

Minimally coupled scalar fields, corresponding to $p=1$, have been
considered in other work \cite{NOO1}.
Let us now discuss the detailed analysis for $p=2$ before considering
the generalisation to $p > 2$. 
Expanding out the Einstein equations in powers of $x$, the leading
order terms determine $\gamma^2$ in terms of $\gamma^0$ as 
\be
[\gamma^2_{ij}]
= \frac{1}{(d-2) l^2} \left [ R^{0}_{ij} - \frac{1}{2(d-1)}
  R^{0} \gamma^{0}_{ij} \right ], \label{gam2}
\ee
where $R^{0}$ and $R^{0}_{ij}$ are the Ricci scalar and Ricci tensor
respectively of the metric $\gamma^0$. For $p \ge 2$,
the $p$-form does not affect the
relationship of $\gamma^2$ to $\gamma^0$ (\ref{gam2})
which is found in the absence of matter fields. 
The $x^2$
term in the $\R_{xx}$ equation of motion gives us the relationship
\be
\tr[(\gamma^{0})^{-1} \gamma^{4}] = \frac{1}{4}  \tr(
(\gamma^{0})^{-1}\gamma^{2} (\gamma^{0})^{-1}\gamma^{2}) + \frac{1}
{16 (d-1) l^2} (F^{0}_2)^2, \label{impl}
\ee
where we contract $F^{0}_2$ again using the leading order metric
$\gamma^0$. Using the $\R_{ij}$ equation of motion at the same order 
we find that 
\be
[\gamma^{4}]^{1}_{ij} = \frac{1}{4 (d-4) l^2} (F^{0}_{2})_{ik}
(F^{0}_{2})^{k}_{\; j} - \frac{3}{16 (d-1) (d-4) l^2} (F^{0}_2)^2
\gamma^{0}_{ij}, \label{one}
\ee
with the other component of $\gamma^4$ being given in terms of the
curvature as the usual expression
\ba
[\gamma^{4}]^{2}_{ij} &=&  
\frac{1}{(d-2)^2(4-d) l^4} \lbrace R^{0}_{il} R^{(0)l}_{j} -
\frac{1}{(d-1)} R^{0} R^{0}_{ij} + \frac{1}{4} (R^{0}_{kl})^2
\gamma^{0}_{ij} \label{gam4} \\
&& \hspace{60mm} - \frac{(d+4)}{4 (d-1)^2} (R^0)^2 \gamma^{0}_{ij}
... \rbrace, \nonumber
\ea
where the ellipses denote terms involving covariant derivatives of the
curvature.
Note that although the expression (\ref{one})
is ill-defined when $d=4$ in this case we will only need to use
the well defined trace (\ref{impl}) to determine divergent terms. 
We can expand out $(F_2)^2$ as
\ba
(F_2)^2 &=& x^4 (F^0_2)^2 + 2 l^2 x^6 (A^1_1)^2 - x^6 (d A^1)_{ij}
(F^0_2)^{ij} + \frac{1}{(d-2)(d-1) l^2} x^6 R^{0} (F^{0}_2)^2 \nonumber \\
&& \hspace{50mm} 
- \frac{2}{(d-2)l^2} x^6 (R^0)^{l}_{j} F^{(0)kj}_{2} F^{0}_{(2)kl}, 
\label{abov}
\ea
where all contractions use $\gamma^0$. Furthermore from the $x^4$ term
in the $\R_{xx}$ equation of motion we derive the relationship 
\be
\tr[ (\gamma^0)^{-1} \gamma^6] = - \frac{1}{3} \tr[ ( (\gamma^0)^{-1}
\gamma^2)^3 ] - \frac{1}{24} (A_1^1)^2 + \frac{1}{48 l^2 (d-1)}
(F_2)^2_{{\cal{O}}(x^6)},
\ee
where the last subscript indicates that we use the coefficient of the
$x^6$ term in (\ref{abov}). Now let us expand the metric; 
as well as terms depending only on $\gamma^0$
and its derivatives 
\ba
\sqrt{g}^{(1)} &=& \frac{\sqrt{\gamma^0}}{l x^{d+1}} \lbrace 1 + \frac{1}{2}
x^2 \tr( (\gamma^{0})^{-1} \gamma^{2})
+ \frac{1}{8}
x^4 [\tr( (\gamma^{0})^{-1} \gamma^{2})]^2  - \frac{1}{8} x^4 
\tr( [(\gamma^{0})^{-1} \gamma^{2}]^2) \label{E1} \\
&& 
- \frac{3}{16} x^6 \tr[ (\gamma^{0})^{-1} \gamma^{2}] 
[\tr( (\gamma^{0})^{-1} \gamma^{2})]^2 + \frac{1}{4} x^6 
\tr( [(\gamma^{0})^{-1} \gamma^{2}]^3) \nonumber \\
&& + \frac{1}{16} x^6 [
\tr( (\gamma^{0})^{-1} \gamma^{2})]^3 - \frac{1}{2} x^6 
\tr[ (\gamma^{0})^{-1} \gamma^{2}(\gamma^{0})^{-1} \gamma^{4}]_{F_2=0}
+ ...  \rbrace. \nonumber
\ea
one has the following terms dependent on the vector field
\ba
\sqrt{g}^{(2)} &=& \frac{\sqrt{\gamma^0}}{l x^{d+1}} \lbrace
\frac{x^4}{32 (d-1) l^2} (F^{0}_2)^2 - \frac{x^6 (1-2d)}{48 (d-1)}
(A_1^1)^2 - \frac{(7d -10) x^6}{96 l^4 (d-1)(d-4)} R^{(0)i}_{j}
F^{(0)jk} F^{0}_{ki} \nonumber \\
&& + \frac{x^6}{96 l^2 (d-1)} (dA^1) \cdot F^{0}_2
 + \frac{(3d-4)}{192 l^4 (d^2-1) (d-4)} R^{0} (F^{0}_2)^2 \rbrace.
\label{E2}
\ea
The on-shell bulk action derived from (\ref{pform}) is
\be
I_{\rm{bulk}} = \frac{1}{8 \pi G_{d+1}} \int_{M} d^{d+1}x \sqrt{g} (d l^2 +
\frac{1}{4(d-1)} F_2^2). 
\ee
Substituting the explicit form for the metric (\ref{E1}) and
(\ref{E2}) and 
integrating we find that 
as well as divergent terms
depending only on $\gamma^0$ and its curvature invariants  
\be
I^{(1)} 
=  - \frac{(d-1)l}{8 \pi G_{d+1} \epsilon^d} \int d^dx \sqrt{\gamma^0}
- \frac{(d-4)(d-1)}{16 \pi (d^2- 4) G_{d+1} l \epsilon^{d-2}} \int 
d^dx \sqrt{\gamma^0} R^{0} + ... \label{usu}
\ee
which can be removed (at least for $d$ odd) by the counterterms given in 
(\ref{ct}), there are additional possible divergences in the bulk
action 
\be
I^{(2)}_{\rm{bulk}} 
=  \frac{1}{8 \pi G_{d+1} l} \int \frac{d^{d+1}x}{x^{d+1}} 
\sqrt{\gamma^0} \lbrace x^4 \frac{(d+8)}{32(d-1)} (F^{0}_2)^2 +
... \rbrace,
\ee
as well as possible divergences in the surface action 
\be
I_{\rm{surf}} = - \frac{1}{8 \pi G_{d+1}} \int d^dx \sqrt{\gamma^0}
\lbrace \frac{d-4}{32(d-1) l \epsilon^{d-4}} (F^0_2)^2 + ... \rbrace.
\ee
There are no divergences for $d < 4$ as previously stated. 
In $d=4$ there is a logarithmic divergence due to the Weyl anomaly
term \cite{HK}
\be
I = - \frac{\ln \epsilon}{64 \pi G_5 l^3} \int d^4x \sqrt{\gamma^{0}}
\left [ (R^{0}_{ij})^2 - \frac{1}{3} (R^0)^2 \right ]. \label{weyl}
\ee
and an additional logarithmic divergence in the action given by 
\be
I_{\rm{log}} = \frac{1}{64 \pi G_{5} l} \ln \epsilon \int d^4 x
\sqrt{\gamma^0} (F^{0}_2)^2. \label{logf}
\ee
Note that the field equation (\ref{feq}) in this case implies that 
$F^0_2$ is both closed and co-closed in the metric $\gamma^0$. 
In $d > 4$ the same term will cause a power law divergence 
in the action
\be
I_{\rm{div}} = - \frac{1}{256 \pi G_{d+1} l \epsilon^{d-4}}
\frac{(d-8)}{(d-4)} \int d^dx \sqrt{\gamma^0} (F^0_2)^2, \label{u1}
\ee
which can be removed by a counterterm of the form 
\be
I_{\rm{ct}} = \frac{1}{256 \pi G_{d+1} l} \int d^dx \sqrt{\gamma}
\frac{(d-8)}{(d-4)} (F_2)^2. \label{ct-4}
\ee
as advertised in \S \ref{two}. 

We should briefly mention another application of these
results. We are concerned here with the definition of local
counterterms which render the action finite as one takes the limit of
$\epsilon \rightarrow 0$, that is, the boundary approaches the true
boundary. In the context of the Randall-Sundrum scenario \cite{RS}, 
however, one would keep the boundary at finite $\epsilon$ as in
\cite{G}. In this case, (\ref{logf}) and (\ref{u1}) would correspond
to part of the conformal field theory action on the brane.

In $d=6$ as well as the logarithmic divergence associated with 
the Weyl anomaly of the dual theory, which is given by 
\ba
I_{\rm{log}} &=&  \frac{\ln \epsilon}{8^4 \pi G_6 l^3} \int d^6x
\sqrt{\gamma^{0}} \lbrace \frac{3}{50} (R^{0})^3 + R^{(0)ij} R^{(0)
  kl} R^0_{ijkl} - \frac{1}{2} R^0 R^{(0)ij}R^{0}_{ij} \\
&& + \frac{1}{5} R^{(0)ij}D^{0}_iD^{0}_j R^0 - \frac{1}{2}R^{(0)ij}
\Box^{0} R^{0}_{ij} + \frac{1}{20} R^{0} \Box^{0} R^{0} \rbrace,
\nonumber 
\ea
as was found in \cite{HK},
we have an anomaly of the form 
\ba
I_{\rm{log}} &=& \frac{1}{8 \pi G_{7} l} \ln \epsilon \int d^6 x
\sqrt{\gamma^0} \left [ \frac{1}{16 l^2} R^{0} (F^{0}_2)^2 - \frac{1}{8l^2}
  R^{(0)ij} (F^{0}_{2})^{\; l}_{i} (F^{0}_{2})_{jl} 
+ \frac{1}{16} (d A^1)^{ij} (F^{0}_2)_{ij} \right ]; \nonumber
\\ &=& \frac{1}{8 \pi G_{7} l^3} \ln \epsilon \int d^6 x
\sqrt{\gamma^0} \lbrace \frac{1}{16} R^{0} (F^{0}_2)^2 - \frac{1}{8}
  R^{(0)ij} (F^{0}_{2})^{\; l}_{i} (F^{0}_{2})_{jl} \label{logf6}
\\
&& \hspace{60mm} 
+ \frac{1}{64} (F^{0}_2)^{ij} \left [ D^{(0)}_{j}D^{(0)k} F^{0}_{ki} - 
D^{(0)}_{i} D^{(0)k} F^{0}_{kj} \right ] \rbrace,  \nonumber
\ea
where in the latter equality we have used the field equation (\ref{feq}).
For $d > 6$, we will need to include an additional counterterm of the
form 
\ba
I_{\rm{ct}} &=& \frac{1}{8 \pi G_{d+1} l^3} \int d^dx \sqrt{\gamma}
\lbrace \frac{(5d-11)}{192 (d-1)^2(d-2)(d-6)} 
R (F_2)^2 + \frac{(7d-66)}{48(d-6)(d-2)} R^{i}_{j} (F_2)_{ik}
(F_2)^{jk} \nonumber \\ 
&& + \frac{(d-8)}{48(d-4)^2} (D_{i} F^{ij}_2)^2  
+ \frac{(d-12)}{196 (d-4) (d-6)} (F_2)^{ij} (D_{j}D^{k} (F_{2})_{ki} - 
D_{i} D^{k} (F_{2})_{kj}) \rbrace.  \label{ct-6}
\ea
The counterterms (\ref{ct-4}) and (\ref{ct-6}) will be adequate for $d <
6$ which includes all gauged supergravity theories of current interest.
For completeness, let us mention that 
it is straightforward to extend the analysis to minimally coupled
$p$-forms of higher order with the following results: the $p$-form 
term in the anomaly for $d =2p$ is given by
\be
I_{\rm{log}} = \frac{1}{32 \pi p G_{d+1} l} \ln \epsilon \int d^{2p}x
\sqrt{\gamma^{0}} (F^0_p)^2,
\ee
whilst for $d > 2p$ we will find a divergence of the form
\be
I_{\rm{div}} = - \frac{1}{64 \pi p^2 G_{d+1} l \epsilon^{d-2p}} 
\frac{(d-4p)}{(d-2p)} \int d^dx (F^0_p)^2,
\ee
which can be removed by the counterterm
\be
I_{\rm{div}} =  \frac{1}{64 \pi p^2 G_{d+1} l } 
\frac{(d-4p)}{(d-2p)} \int d^dx (F_p)^2.
\ee
There will of course be additional terms in the anomaly for $d = 2p +2n$
depending on derivatives of $F_p$ and the curvature. 

\subsection{Magnetic strings in five dimensions}
\noindent

Solutions of gauged supergravity theories in
which these anomalies and counterterms play a role have been constructed. 
The simplest example is the magnetic string
solution of cosmological Einstein-Maxwell theory in five dimensions
\cite{CS}, \cite{KS} which is given by 
\ba
ds^2 &=& (l r)^{\frac{1}{2}} (lr + \frac{1}{3 lr})^{\frac{3}{2}}
(d\tau^2 + dz^2) + \frac{dr^2}{(lr + \frac{1}{3 lr})^2} + r^2
(d\theta^2 + \sin^2 \theta d\phi^2); \nonumber \\
F_2 &=& - \frac{1}{\sqrt{3} l} \sin \theta d\theta \wedge d\phi.
\ea
Note that the magnetic charge is quantised in units depending on the
cosmological constant \cite{LR}. Like the BPS electric black hole
solutions in four and five dimensional gauged supergravity theories
the magnetic string solution represents a naked
singularity. Substituting the fields into the bulk action, 
we find that the potential logarithmic divergence is in fact absent.
Note that in calculating the bulk action we have
to introduce a UV cutoff around the singularity for the BPS solution but
this won't affect the determination of the IR divergences.

To check on this result, let us
calculate the logarithmic terms in the action directly, using
the boundary metric $\gamma^0$ 
\be
\gamma^0_{ij} dx^{i} dx^{j} =  l^2 (d\tau^2 + dz^2) + 
(d\theta^2 + \sin^2 \theta d\phi^2).
\ee
The dual conformal theory hence has a background of $R^2 \times S^2$. 
Then the Weyl anomaly (\ref{weyl}) is given by 
\be
I_{\rm{log}} = - \frac{\ln \epsilon}{96 \pi G_{5} l^3} \int d^4x
\sqrt{\gamma^{0}}. \label{log}
\ee
However, substituting into the expression (\ref{log}), we find an
equal and opposite 
contribution from the vector field, giving zero total 
divergence. 
We should mention here that the logarithmic term 
for the ``dual'' electric black holes in five dimensions vanishes
\cite{EJM}; this
is because the Weyl anomaly for a field theory on $R^1 \times S^3$
vanishes. However, both anomaly cancellations appear to be accidental.

\subsection{Magnetic three-branes in seven dimensions}
\noindent

Cosmological Einstein-Maxwell gravity in seven dimensions admits magnetic
three-brane solutions of the form 
\ba
ds^2 &=& (l r)^{2} (1 + \frac{1}{5 l^2 r^2})^{\frac{5}{4}}
(d\tau^2 + ds^{2}(E^3)) + \frac{dr^2}{(lr + \frac{1}{5 lr})^2} + r^2
(d\theta^2 + \sin^2 \theta d\phi^2); \nonumber \\
F_2 &=& - \frac{2}{\sqrt{5} l} \sin \theta d\theta \wedge d\phi.
\ea
This solution is obviously very closely related to the magnetic string
solution in five dimensions; however, it is not a solution
of the gauged supergravity theories which arise in seven dimensions 
from compactification of eleven-dimensional supergravity on a sphere,
since the latter does not admit an Einstein-Maxwell
truncation. Calculating the bulk action, we find that again the
logarithmic divergence does not appear.
Calculation of the
logarithmic divergence both directly and using the expression 
(\ref{logf6}) allows us to check the coefficients in
(\ref{logf6}). Again there appears to be no profound reason why the
total logarithmic anomaly should vanish.  

\section{Scalar fields with gauged supergravity potentials}
\label{four} 
\noindent

The action for a scalar field with potential is 
\be
I_{\rm{bulk}} = - \frac{1}{16 \pi G_{d+1}} \int_{M} d^{d+1}x \sqrt{g} 
\left [ \R + l^2 V(\phi) - \frac{1}{2} (\partial \phi)^2 \right ], 
\label{scal}
\ee
where
the equations of motion derived from the action are
\ba
\R_{mn} &=& \frac{1}{2} (\partial_{m} \phi) (\partial_{n} \phi) -
\frac{1}{(d-1)} l^2 V(\phi) g_{mn}; \nonumber \\
\R &=& \frac{1}{2} (\partial \phi)^2 - \frac{(d+1)}{(d-1)} l^2 V(\phi); \\
D_{m} \partial^{m} \phi &=& - l^2 (\frac{\partial V}{\partial
  \phi}). \nonumber 
\ea
The leading order terms in the Einstein equations imply that if the
metric behaves as (\ref{asym}) near the conformal boundary the scalar
field must tend to a value on $N$, $\phi \rightarrow \phi^{0}(x^i)$,
such that the scalar potential takes the constant value $V(\phi^{0}(x^i)) =
d(d-1)$. Furthermore, using the leading order term in the 
dilaton equation of motion, we find that 
\be
\left (\frac{\partial V}{\partial \phi} \right )_{\phi = \phi^0(x^i)} = 0.
\ee
In the vicinity of the conformal boundary the scalar field
must behave as 
\be
\phi(x, x^{i}) = \phi^{0} + x^2 \phi^{2}(x^i) + x^4 \phi^{4}(x^i)......,
\ee
and the potential may be expanded about its boundary value as
\ba
V(\phi) &=& d(d-1) + \frac{1}{2} V^2(x^i) (\phi - \phi^0)^2 +
\frac{1}{6} V^3(x^i) (\phi - \phi^0)^2 + ...\\
&=& d(d-1) + \frac{1}{2} x^2 V^2(x^i) (\phi^1(x^i))^2 + x^3
(V^2(x^i) \phi^1(x^i) \phi^2(x^i) + \frac{1}{6} V^3(x^i)
(\phi^1(x^i))^3) + ..., \nonumber
\ea
where $V^{i}(x^i)$ is the $i^{\rm{th}}$ derivative of $V$ with respect to
$\phi$ evaluated at $\phi^0$. Then the derivative of the potential 
can be expanded as
\be
(\frac{\partial V}{\partial \phi}) = x V^2(x^i) \phi^1(x^i) + x^2
(V^2(x^i) \phi^2(x^i) + \frac{1}{2} V^3(x^i) (\phi^1(x^i))^2) + ....
\ee
So far, we have taken the fields $\phi^0$ and $V^{i}(x^i)$ to be arbitrary, 
but of course for the
potentials which arise in maximal supergravities  
arbitrary values of the fields cannot be chosen; in particular, what is 
relevant here is that the second derivative of the potential at the boundary 
is fixed. 

So let us consider here scalar fields in $D$-dimensional maximal supergravity; 
these parametrise the coset $E_{11-D}/K$ where $E_{n}$ is the exceptional 
group and $K$ is its maximal compact subgroup. Focusing on the $SL(N,R)$
subgroup of $E_{n}$ and using the local $SO(N)$ transformations to 
diagonalise the scalar potential we are led to the form \cite{CGLP}
\be
V = \frac{d(d-1)}{N(N-2)} \left ( (\sum_{i=1}^{N} X_i)^2 - 2 
(\sum_{i=1}^{N} X_i^2) \right ), \label{pot}
\ee
where $d = D-1$ and
the $N$ scalars $X_i$ can be parametrised in terms of $(N-1)$ 
independent scalars $\phi_{\alpha}$ as
\be
X_{i} = e^{-\frac{1}{2} b_{i}^{\alpha} \phi_{\alpha}},
\ee
where the $b_{i}^{\alpha}$ are the weight vectors of the fundamental 
representation of $SL(N;R)$ normalised so that 
\be
b_{i}^{\alpha} b_{j}^{\alpha} = 8 \delta_{ij} - \frac{8}{N}, \hspace{10mm} 
\sum b_{i}^{\alpha} = 0.
\ee
Then the potential (\ref{pot}) has a minimum at $X_{i} = 1$ for $N \ge 5$ 
(which includes all cases considered here), at which point $\phi_{\alpha} =0$ 
and $V = d(d-1)$. Explicitly differentiating we find that the second 
derivatives at this minimum are given by 
\be
\frac{\partial^2 V}{\partial \phi_{\alpha} \partial \phi_{\beta}} 
 = \frac{d(d-1)}{N(N-2)} b_{i}^{\alpha} b_{i}^{\beta}.
\ee
Using the properties of the weight vectors we find that 
\be
b_{i}^{\alpha} b_{i}^{\beta} = 4 (N-4) \delta^{\alpha \beta} 
\hspace{10mm} \forall \alpha.
\ee
We will be interested in maximal supergravities in $D=4,5,7$ for which 
$N = 8,6,5$ respectively. Substituting in these values, we find that the 
second derivatives of the potentials are given by 
\be
\frac{\partial^2 V}{\partial \phi_{\alpha} \partial \phi_{\beta}} 
 = 2 (d-2) \delta^{\alpha \beta}. \label{der2}
\ee
The same expression is found for the non-maximal supergravity potential in 
six dimensions discussed in \cite{CGLP}; this potential arises
naturally from the reduction of massive IIA supergravity on $S^4$.
The significance of (\ref{der2}) is the following; since the 
form of the potential fixes the asymptotic values of the scalar fields to be
zero, each scalar field can be expanded as
\be
\phi(x,x^i) = x^{k} \phi^{k}(x^i) + x^{k+1} \phi^{k+1}(x^i) + ...
\ee
The leading order term in the scalar equation is given by
\be
\left [ k (d - k) - V^2 \right ] \phi^k(x^i) = 0,
\ee
and so using (\ref{der2}) we find that $k = d-2$. Note that it is the 
specific form of the potential which forces the scalar to behave
as $x^{d-2}$ at infinity; in a more general potential, we might have  
leading order terms with $k < d-2 $ 
which would give rise to further divergences in the action. 
The form of the 
potential effectively ensures that the scalar charge is finite; it can be 
expanded as 
\be
V(\phi) = d(d-1) + (d-2) x^{2(d-2)} (\phi^{d-2}(x^{i}))^2 + ...
\ee
Since the on-shell 
bulk term in the Einstein action is given by
\be
I_{\rm{bulk}} = \frac{l^2}{8 \pi G_{d+1} (d-1)} \int_{M} d^{d+1}x
\sqrt{g} V(\phi),
\ee
using just the asymptotic form of the metric (\ref{asym}) 
the only possible scalar field
divergences are in $d = 3,4$. In the former case, the Einstein equations 
give us the relationship
\be
\tr [(\gamma^0)^{-1} \gamma^2 ] = \frac{R^{0}}{4l^2} - \frac{3}{8} 
(\phi^1(x^i) )^2,
\ee
which is enough to determine the dependence of the divergence on the scalar 
fields
\be
I_{\rm{div}} = - \frac{5l}{256 \pi G_{4} \epsilon} \int d^3x \sqrt{\gamma^0}
(\phi^1(x^i))^2.
\ee
This means that one needs an additional counterterm in the action of the
form 
\be
I_{\rm{ct}} =  \frac{5l}{256 \pi G_{4}} \int d^3x \sqrt{\gamma}
(\phi)^2.
\ee
One will need this term to regularise the action of the
four-dimensional charged black holes discussed in \cite{CK}, \cite{K},
\cite{DL}, \cite{S}.
For $d=4$, $\gamma^2$ is unaffected by the scalar fields, but 
\be
\tr [ (\gamma^0)^{-1} \gamma^4] = - \frac{1}{3} (\phi^2)^2,
\ee
which is enough to determine that 
\be
I_{\rm{div}} = \frac{l}{48 \pi G_5} \ln \epsilon \int d^4 x \sqrt{\gamma^0} 
(V^2 -4) (\phi^2(x^i))^2 = 0.
\ee
That is, the potential anomaly term in $d=4$ is absent. We should be 
unsurprised by this for two reasons. The first is that the
thermodynamics of charged black holes in five dimensions \cite{BCS},
\cite{BCS1} has been
extensively discussed in the context of stability analysis and no
logarithmic term in the action was found. Secondly, 
there is no candidate for an anomaly of this form in the dual
conformal field theory. 

\section{Gauged supergravity in seven dimensions} \label{foura}
\noindent

The analysis of the two previous sections assumes that one can
consistently truncate a supergravity theory to an Einstein-Maxwell
theory or to an Einstein-dilaton theory. However, this isn't generally
the case. Although solutions of Einstein-Maxwell theory are consistent
solutions of four and five dimensional gauged supergravities, they are
not solutions of gauged supergravity in seven dimensions. 

Let us consider the following truncation of $D=7$ gauged supergravity,
containing gravity, two scalar fields $\phi_1,\phi_2$, two vector fields
$F_1,F_2$ and a single three-form potential $C$ discussed in \cite{LM}
\ba
I_{\rm{bulk}} &=& 
\frac{1}{16 \pi G_{7}} \int_{M} d^7x \sqrt{g} 
\lbrace R + 2 l^2 V(\phi_1,\phi_2) - 5 (\partial(\phi_1+\phi_2) )^2 - 
(\partial(\phi_1-\phi_2) )^2 \nonumber \\
&& \hspace{30mm} - e^{-4 \phi_1} (F_{1})^2 
 e^{-4 \phi_2} (F_{2})^2 + 4 l^2 e^{-4 \phi_1 - 4 \phi_2} C^2 \\
&& \hspace{30mm} -
\frac{l}{3} \epsilon^{mnpqrst} C_{mnp} \partial_{q} C_{rst} +
\frac{1}{\sqrt{3}} \epsilon^{mnpqrst} C_{mnp} F_{(1)qr} F_{(2)st} +
... \rbrace, \nonumber
\ea
where the ellipses denote terms that are not relevant here. We have
simplified the field content by focusing on a $U(1)^2$ truncation of
the gauge group and by diagonalising the potential. The resulting scalar
potential is given by 
\be
V(\phi_1,\phi_2) = 8 e^{2 \phi_1 + 2 \phi_2} + 4 e^{-2 \phi_1 - 4
  \phi_2} + 4 e^{4 \phi_1 + 2 \phi_2} + e^{-8 \phi_1 - 8 \phi_2},
\ee
which is of the form discussed in the previous section and the
field equations are 
\ba
D_{m}D^{m}(3 \phi_1 + 2 \phi_2) &=& - e^{-4 \phi_1}(F_1)^2 + 4 l^2
e^{-4  \phi_1 -4 \phi_2} C^2 - \frac{1}{2} \frac{\partial V}{\partial
  \phi_1}; \nonumber \\
D_{m}D^{m}(3 \phi_2 + 2 \phi_1) &=& - e^{-4 \phi_2}(F_2)^2 + 4 l^2
e^{-4  \phi_1 -4 \phi_2} C^2 - \frac{1}{2} \frac{\partial V}{\partial
  \phi_2}; \nonumber \\
D_{m}(e^{-4\phi_1} F_{1}^{mn}) &=& \frac{1}{2\sqrt{3}}
\epsilon^{mnpqrst} D_{m}(F_{(2)pq} C_{rst}); \label{eaa} \\
D_{m}(e^{-4\phi_2} F_{2}^{mn}) &=& \frac{1}{2\sqrt{3}}
\epsilon^{mnpqrst} D_{m}(F_{(1)pq} C_{rst}); \nonumber \\
e^{-4 \phi_1 -4 \phi_2} C_{mnp} &=& \frac{1}{12 l} \epsilon_{mnp}^{qrst}
\partial_{q} C_{rst} - \frac{1}{8 \sqrt{3} l^2} \epsilon_{mnp}^{qrst}
F_{(1)qr} F_{(2) st}. \nonumber
\ea
The key point is that one cannot consistently set the scalars and the
three-form potential to zero when the vector fields are excited. If
the wedge product of the two vector field strengths vanishes, 
then one can set the three-form to zero, as for the known electric 
black hole solutions \cite{LM}, but even this is not possible in general. 

\bigskip

A consistent further truncation is to set $F_2 =0$, $C=0$ and $2 \phi_1 + 3
\phi_2 = 0$, in which case the field equations become
\ba
D_{m}D^{m} \phi &=& -\frac{3}{5} e^{-4\phi} F^2 - \frac{12}{5} l^2
(e^{\frac{2\phi}{3}} - e^{-\frac{8 \phi}{3}} ); \hspace{5mm}
D_{m} (e^{-4\phi} F^{mn}) = 0; \label{seq} \\
R_{mn} &=& -\frac{2 l^2}{5} \left ( 12 e^{\frac{2 \phi}{3}} + 3
  e^{\frac{8 \phi}{3}} \right ) + \frac{10}{3} \partial_m \phi
\partial_n \phi + 2 e^{-4 \phi} \left ( F_{mp} F_{n}^{p} -
  \frac{1}{10} F^2 g_{mn} \right ), \nonumber
\ea
where we have dropped the subscripts on the fields for notational
simplicity. Following the arguments of the previous section, using the
leading order form of the asymptotic metric and the Einstein equations
fixes the expansion of the fields about the boundary to be
\ba
\phi(x,x^{i}) &=& x^{k} \phi^{k}(x^{i}) + x^{k+2} \phi^{k+2} + ...; \\
F &=& F^{0}_2 + x dx \wedge A^1_1 + x^2 F^2_1 + x^2 dx \wedge A^2_1 +
....., \nonumber
\ea
where $k > 0$ and we use the same notation for the vector field as in
\S \ref{two}. Now substituting into the scalar field equation
(\ref{seq}) we get the following expression
\be
l^2 (k^2 - 6k + 8) x^k \phi^k(x^i) + \frac{3}{5} x^4 (F^0_2)^2 =
{\cal{O}}(x^{k+2},x^6),
\ee
where as usual we raise and lower indices in the leading order induced
metric $\gamma^0$. 
The only way to balance leading order terms is to take $k = 4$ and  $\phi^4
\neq 0$ (as found in the previous section) {\it but} this forces $F^0_2
= 0$. 

Hence a finite vector field is not induced on the boundary;
one cannot have a non-zero magnetic charge in the
bulk with this ansatz. Furthermore, 
since $F^0_2$ vanishes, as previously mentioned in \S \ref{two}
the Maxwell equations fix the leading
order behaviour of $F^2$ to be $x^8$ or smaller near the
boundary, and thus the vector field will not contribute to any 
divergences of the action. 

This result should be independent of the particular truncation we have
taken. Analysing (\ref{eaa}) one sees that the scalar field
equations will always force the magnetic charge in the Abelian
truncation of the theory to vanish, given the
form of the scalar potential. One would also expect this obstruction to
persist in the full non-Abelian theory. 

\bigskip

Another consistent truncation is to set $F_1 = F_2 = 0$ and $\phi_1 =
\phi_2$; this is relevant if we are trying to induce finite magnetic
charge with respect to the three form potential. 
The relevant field equations are then
\ba
D_{m} D^{m} \phi &=& \frac{4}{5} e^{-8 \phi} C^2 - \frac{8}{5} \left ( 2
  e^{4 \phi} + 3 e^{- 6 \phi} - e^{-16 \phi} \right ); \label{upt} \\
e^{-8 \phi} C_{mnp} &=& \frac{1}{12 l} 
\epsilon_{mnp}^{qrst} \partial_{q} C_{rst}. \nonumber
\ea 
In this case it is the field equation for the three form which is
important; the scalar field equation would not prohibit finite 
magnetic three-form charge. 
Suppose we look for a leading order
expansion of $C$ of the form
\be
C = x^a C^{a}_3 + x^b dx \wedge A^{b}_2 + ...
\ee
where $C^a_3$ and $A^b_2$ are three and two forms respectively,
defined on the induced six-dimensional hypersurface.
Then the self-duality equation (\ref{upt}) 
fixes $a=2$ and $C^2_3$ to be self-dual
in the six-dimensional induced metric $\gamma^0$, in agreement with
the discussions of 
\cite{LM}. Furthermore, the next order terms in this equation fix $b =
3$ with 
\be
A^3_2 = \frac{2}{l^2} \ast dC^{2}_{3}, \label{cond}
\ee
where the dual is again taken in the metric $\gamma^0$. This means that the
leading order contribution to the action from terms in $C$ is $x^{10}$,
and hence there are no IR divergences resulting from the inclusion of
$C$ terms. 
Physically, this means that one cannot find solutions of the gauged 
supergravity with an asymptotic metric of the form 
(\ref{asym}) which have finite magnetic three-form charge.  

\bigskip

For completeness, let us also consider the truncation with 
$\phi_1 = \phi_2 =0$ and $F_1 = F_2$. 
Consistency of the truncation requires that 
\be
F^2 = 4 l^2 C^2; \hspace{10mm} D_{m} F^{mn} = \frac{1}{2 \sqrt{3}}
\epsilon^{mnpqrst} D_{m}(F_{pq} C_{rst});
\ee
and we also have to satisfy the last field equation in (\ref{eaa}).
Together these conditions prove to be very restrictive, 
and the leading order behaviour of $C$ and $F$ is fixed to be 
\be
C = x^2 C^2_3 + x^3 dx \wedge A^3_2 + ...; \hspace{10mm} 
F = x^3 dx \wedge A^3_1 + ....,
\ee
where $C^2_3$ is self-dual in $\gamma^0$ and $A^3_2$ satisfies the
condition (\ref{cond}) as before. $A^3_1$ satisfies the gauge
condition $D_{m} (A^3_1)^m = 0$ and satisfies the relation $(A^3_1)^2
= 2 (C^2_3)^2$. As in the above truncations, no finite magnetic fluxes - and
hence no anomalies - are induced in the boundary field theory. 

\section{Anomalies in conformal field theories} \label{five}
\noindent

The general structure of the gravitational action for $d$ even is
\be
I = \frac{1}{16 \pi G_{d+1}} \int d^d x
\left [\sqrt{\gamma^0} \left ( a^{0} \epsilon^{-d} + ... + a^{d-2}
  \epsilon^{-2} + a^{d} \ln \epsilon \right ) \right]  + {I}_{\rm{fin}},
\ee
where ${I}_{\rm{fin}}$ is finite in the $\epsilon \rightarrow 0$
limit. After subtraction of the divergent counterterms, including the
logarithmic term, we are left with a renormalised effective action
with a finite limit as $\epsilon \rightarrow 0$. Its variation under a
conformal transformation $\delta \gamma^{0} = 2 \delta \lambda
\gamma^{0}$ is of the form 
\be
\delta {I}_{\rm{fin}} =  \int d^dx \sqrt{\gamma^0} {\cal{A}} \delta
\lambda,
\ee
where the anomaly ${\cal{A}}$ is given by
\be
{\cal{A}} = \frac{a^d}{8 \pi G_{d+1}}.
\ee
Hence as first discussed in \cite{HK} one can relate the logarithmic
divergence of the gravitational action to Weyl anomalies of the conformal
field theory. On general grounds \cite{DS}, \cite{OP}, 
the gravitational part of the coefficient that
appears in the anomaly must be of the form
\be
a^{d} = d l^{1-d} (E^d + I^d + D^{(0)i}J^{d-1}_{i}), \label{gf}
\ee
where $E^{d}$ is proportional to the d-dimensional Euler density (the
type A anomaly) and $I^d$ is a conformal invariant (type B anomaly). 
The current term depending on $J^{d-1}$ 
is trivial in the sense that it can be obtained by
variation of covariant counterterms. 

\bigskip

Anomalies which combine Weyl invariance with other symmetries can be
analysed in much the same way as pure Weyl anomalies. Consider vector
field sources in $d=4$; using the duality relations $G_5 = 8 \pi^3 l^3
g_{s}^2$ and $l = (4 \pi g_s N)^{-\frac{1}{4}}$ we find that the
anomaly is
\be
{\cal{A}}_{F^0_2} = \frac{1}{256 \pi^4 g_{s}^2 l^6} (F^0_2)^2 =
\frac{N^2}{16 \pi^2} f^2, \label{sgr}
\ee
where in the latter equality we have used the fact that (magnetic) fields
in the bulk behave as $1/l$ \cite{LR} to define $f = l F^0_2$ as the more 
natural field in the dual theory. One expects 
the anomaly in the correlator of the stress tensor
and two vector currents for any conformal theory in 
$d=4$ to be proportional to the Maxwell
Lagrangian density \cite{CJ}, \cite{CE};   
in $d=4$ the only Weyl invariant term of the right
dimension involving $f$ is the Maxwell Lagrangian density. For the
${\cal N} = 4$ $SU(N)$ Yang-Mills theory, following \cite{HK}, 
the anomaly should be
\be
(N^2 -1) \left [6 \alpha_s + 2 \alpha_f + \alpha_v \right ] f^2, \label{ym}
\ee
where the coefficients $\alpha_s$, $\alpha_f$ and
$\alpha_v$ are for non-interacting
scalars, fermions and vectors respectively and we
have included the multiplet of six scalars, two fermions and one vector.
The factor of $(N^2 -1)$ derives as usual 
from the fields taking values in the adjoint of $SU(N)$. We
assume that, as for the Weyl anomaly, there is a renormalisation
theorem which protects the coefficients so that we can ignore
interactions.  Now the coefficients $\alpha_{i}$ are
well-known; indeed they were effectively calculated in the 
original papers \cite{CJ} and \cite{CE} (see also \cite{DDI} and
\cite{CDJ}). Using the explicit values
\be
\alpha_s = \frac{1}{256 \pi^2}; \alpha_f = \frac{1}{64 \pi^2};
\alpha_v = \frac{1}{128 \pi^2},
\ee
we find that the Yang-Mills anomaly (\ref{ym}) 
in the large $N$ limit agrees with what we found in 
the supergravity theory (\ref{sgr}) even though we have ignored
interactions! Such an agreement is perhaps less surprising given that
it is already known that the pure Weyl anomaly is not renormalised 
\cite{HK}; the latter should be related to the vector anomaly by
supersymmetry. 

\bigskip

Now let us consider vector fields in $d=6$. Since there is a
non-vanishing logarithmic divergence in the action 
when one includes magnetic vector fields in cosmological
Einstein-Maxwell theory, we expect that in the (unknown) dual
conformal field theory there should be an anomaly in the correlator of
the stress-energy tensor and two vector currents. 
In analogy to the discussions of
the Weyl anomaly, we need to construct a basis of 
Weyl invariant polynomials involving the curvature and the vector
field. Appropriate polynomials will be constructed from one curvature
tensor and two Maxwell fields, or from two derivatives and two Maxwell
fields. A basis of possible contractions is given by 
\ba
\left[ V_{a} \right ] &=& \lbrace R f_{ij}f^{ij}, 
R^{ij} f_{jl} f_{i}^{\; l}, (D^i f_{ij})^2, 
f^{ij} D_{ [i} D_{|k|} f^{\; k}_{j]}, 
\Box (f^{ij} f_{ij}),  D_{i}D_{j} f^{i}_{\; k} f^{kj},  \\
&& \hspace{20mm}(D_{i}f^{jk})(D^{i}f_{jk}),
f^{ij} D_{k} D_{[i} f^{\; k}_{j]}, f^{ij} \Box f_{ij}, 
(D_{i}f^{jk})(D^{i}f_{jk})
\rbrace, \nonumber
\ea
where we have rescaled the vector field as above. Now there is a
conformal invariant given by 
\be
I = 2 V_1 - 4 V_2 + V_3 + 2 V_4 - 2 V_5 + 2 V_6;
\ee
one should be able to construct other conformal invariants from
$\lbrace V_a \rbrace$ but they do not appear in the anomaly.
Note also that we can write as derivatives of currents
\be
D_{i} J^{i}_1 = V_3 + V_4; \hspace{10mm} D_{i}J^{i}_{2} = V_5; 
\hspace{10mm} D_{i}J^{i}_{3} = V_6.
\ee
We then find that the anomaly is given by 
\be
{\cal{A}}_{f} = \frac{1}{256 \pi l^5 G_7} \left ( I - D_{i} J^{i}_1 + 2
  D_{i}J^{i}_{2} - 2 D_{i}J^{i}_{3}  \right),
\ee
in accordance with the general anomaly form (\ref{gf}). 

\bigskip

Finally, let us interpret the results of \S \ref{foura} in terms of
the dual conformal field theory. Gauged supergravity in a seven
dimensional anti-de Sitter background should be dual to a boundary
field theory consisting of $(0,2)$ tensor multiplets coupled to
$(0,2)$ conformal supergravity in six dimensions. The boundary values
of the bulk fields are in one-to-one correspondence with the fields of
(off-shell) conformal supergravity defined at the boundary. 

The coupling of a single $(0,2)$ tensor multiplet to conformal
supergravity in six dimensions was discussed in \cite{BSV}. Given the
known result for the Weyl anomaly, one could determine the
vector and three-form anomalies 
using supersymmetry. One could also try to determine
these results directly by analysing a single $(0,2)$ tensor multiplet as
was done for the Weyl anomaly in \cite{T} 
(although this won't give the right answer since it
doesn't for the Weyl anomaly). 
One should find that the predicted anomaly involves the 
invariants given above; there is no reason a priori why it should vanish. 

However, what we have argued in \S \ref{foura} is that
the bulk equations of motion prevent us from inducing finite vector or
three form fields on the boundary. That means that the bulk theory
in this case can only tell us about tensor multiplets in backgrounds
without vector or three form currents switched on. 

\acknowledgements

Financial support for this work was provided by St John's College,
Cambridge. I am grateful for various comments made to me by Arkady
Tseytlin.

\end{document}